\documentclass[twocolumn]{oja}
\usepackage{graphicx}
\usepackage{amsmath}
\usepackage{bm}
\usepackage{hyperref}
\usepackage{orcidlink}
\usepackage{lastpage}
\usepackage{multirow}

\defcitealias{pearce2022}{TP22}

\def\Mcompanion{M_c}
\def\nxa{\textit{Exoplanet Archive}} 
\def\gaia{\textit{Gaia}}

\def\otypes{{\texttt{otypes}}}
\def\ruwe{{\texttt{ruwe}}}
\def\pmra{{\texttt{pmra}}}
\def\pmdec{{\texttt{pmdec}}}
\def\pmraerr{{\texttt{pmra\_error}}}
\def\pmdecerr{{\texttt{pmdec\_error}}}
\def\parallax{\texttt{parallax}}
\def\paxerr{\texttt{parallax\_error}}
\def\corrcoeff#1#2{\texttt{{#1}\_{#2}\_corr}}
\def\bprp{\texttt{bp\_rp}}
\def\vpu{\texttt{visibility\_periods\_used}}
\def\sysnum{\texttt{sy\_snum}}
\def\photgmeanmag{\texttt{phot\_g\_mean\_mag}}
\def\nss{\texttt{non\_single\_star}}

\def\mjup{\text{M$_{\text{J}}$}}
\def\msun{\text{M$_\odot$}}
\def\rsun{\text{R$_\odot$}}

\def\mmin{M_\text{min}}
\def\permax{T_\text{max}}

\def\probpl{P_\text{pl}}
\def\au{{au}}

\begin{document}

\title{An astrometric search for planets in debris disk systems}

\author{Elisabeth M. Penderghast$^1$\orcidlink{0009-0007-6715-197X}}   
\author{Benjamin C. Bromley$^1$\orcidlink{0000-0001-7558-343X}}
\author{Scott J. Kenyon$^2$\orcidlink{0000-0003-0214-609X}}
\author{Joan R. Najita$^3$\orcidlink{0000-0002-5758-150X}}
\affiliation{$^1$Department of Physics and Astronomy, University of Utah, 115 S 1400 E, Salt Lake City, UT, 84112, USA}
\affiliation{$^2$Smithsonian Astrophysical Observatory, 60 Garden Street, Cambridge, MA 02138, USA} 
\affiliation{$^3$NSF’s NOIRLab, 950 N. Cherry Avenue, Tucson, AZ 85719, USA}

\begin{abstract}
Debris disks are likely created and sculpted by planetary bodies in the orbital space they share. The properties of these disks, including mass, orbital extent, and morphology, can be indicators of their planetary shepherds. Recently, T.~D.~Pearce and collaborators placed limits on the masses and orbits of hypothetical planets around 178 stars with debris disks. Here, we consider 176 of these stars, all objects that have astrometric data in the \gaia\ Data Release 3 archive, to assess planet detection from astrometry. Our analysis begins with a set of stellar hosts of known exoplanets, selected to have parallax, apparent magnitude, and color similar to the 176 debris disk systems. We confirm that \gaia's \ruwe\ parameter, a measure of the quality of astrometric fitting to a linear drift model, is sensitive to the presence of massive companions, even planetary ones. We also train a machine-learning model with a range of planet host properties from \gaia, considering more detailed astrometric data beyond \ruwe\ that might indicate orbital motion. In addition, the model incorporates \gaia\ photometry to inform how brightness and stellar type affect detectability, with the risk that selection bias in the training set may affect reliability. Guided by \ruwe\ and predictions of the machine-learning model, we identify stars with debris disks that may host as-yet-undiscovered planets. Overall, the model predictions, \ruwe, and the mass limits from debris disk morphology do not show strong trends. Nonetheless, our top candidates with high \ruwe\ and high planethood probability from the machine learning model will be compelling subjects for time‑series analyses with \gaia\ Data Release~4.
\end{abstract}


\section{Introduction}

Debris disks are powerful tracers of the structure and evolution of planetary systems \citep{wyatt1999, wyatt2008, krivov2010,  hughes2018, manara2023}. Composed of sub-micron to millimeter-sized dust grains and larger, unseen planetesimals, debris disks likely arise from collisions driven by the gravity of planets or planetesimals \citep[e.g.,][]{wyatt2002,kb2002a, kb2002b,dom2003, wyatt2008, mustill2009, kb2010, raymond2011, matthews2014, najita2022}. Although the more massive bodies may be difficult to detect, dust reprocesses and scatters starlight efficiently. Vega hosts the first known debris disk \citep{Aumann1984}. {\it Spitzer} and {\it JWST} images show a smooth, extended structure and a Kuiper-belt analog, yet the system has no planets as massive as Saturn beyond a few \au\ \citep{su2005,su2013,su2024}. But for the dust, this architecture would be missed.

From theoretical arguments \citep[e.g.,][]{kb2001, mustill2009}, the origin of circumstellar debris is gravitational stirring of planetesimals by embedded or adjacent planets. Fragments from gravity-driven collisions between larger bodies create a plethora of smaller ones, which in turn can collide and shatter in a collisional cascade. The resulting dusty debris disk may be bright in reflected light or reprocessed starlight as compared to any planets in the same system. In this way, they become indirect indicators of planetary bodies \citep[e.g.][]{kb2002signpost}. Massive planets tend to carve gaps in debris disks; the presence of a gap might indicate a planet \citep[e.g.][]{greaves1998}. Other evidence of a planet's gravitational influence, including disk asymmetries and warping, can provide stronger cases for their existence \citep[e.g.,][]{wyatt1999}.

Prominent examples of debris disk systems include $\beta$ Pictoris with its warped debris disk between and beyond the host star's known planets \citep{mouillet1997, okamoto2004, lagrange2009, lagrange2019, nowak2020}, and TWA 7, with dynamically cold asymmetric dusty debris \citep{choquet2016, olofsson2018} that revealed one confirmed Saturn-mass planet \citep{lagrange2009} and suggests the presence of another \citep{crotts2025, lacquement2026}. Both $\beta$ Pic and TWA~7 show clear connections between their debris disks and planets.

\citet[hereafter \citetalias{pearce2022}]{pearce2022} leverage the connection between debris disks and the planets that may sculpt them to constrain planetary properties from disk geometry \citep[see also][]{faber2007, mustill2012, pearce2014, morrison2015, nesvold2015, shannon2016, lazzoni2018, regally2018}. \citetalias{pearce2022} examine 178 cold debris disk systems focusing on dust at $\sim$10~AU or further, detected through resolved images from the Atacama Large Millimeter Array and the Herschel Space Observatory, or through spectral energy distribution fitting. Then \citetalias{pearce2022} consider whether disk self-stirring, an interior planet, or multiple planets could explain the morphology of the observed debris disks.

To quantify their assessment, \citetalias{pearce2022} consider theoretical gravitational stirring mechanisms, needed to maintain small grains that would otherwise be cleared out from interaction with starlight, and sculpting by planets whose gravity can carve a disk's inner edge. They show that self-stirring within a disk \citep[e.g.,][]{kb2001, kennedy2010, krivov2018} typically requires an untenably large disk mass to account for the observed debris-disk systems \citep[e.g.,][cf.~\citealt{nkb2022}]{krivov2021}. Multiple small planets interior to a disk can efficiently sculpt the disk's inner edge \citep{faber2007, su2013}, although they cannot always stir the disk, especially if it is extended \citep{mustill2009}. \citetalias{pearce2022} find that models with a single massive planet generally provide both sustained stirring and sculpting \citep[see also][]{pearce2014}. They created a catalog of predicted planets, including the minimum mass and maximum orbital distance that these planets must have if they are responsible for generating the observed dusty debris.
 
Here we explore an independent approach to inferring the presence of the planetary candidates in the 178 debris disk systems from \citetalias{pearce2022}. All but two stars in this catalog are also in the Gaia DR3 archive; Gaia's astronomical measures may be sensitive to the stars' reflex motion arising from a massive companion planet. A strategy of using Gaia's astrometric quality indicators to identify companions has proved successful in a range of contexts, including binary stars \citep{brandt2021, belokurov2020, fabricius2021, kervella2022, gaiabin2023}, substellar objects,  planets \citep[e.g.,][]{holl2023, stefansson2025, vioque2026} and black holes \citep{elbadry2023, chakrabarti2023, muller-horn2025}. On the eve of the release of Gaia Data Release 4 \citep[e.g.,][]{brown2025}, our hope is to identify the most promising candidates for astrometric confirmation in sources for which there is already compelling evidence for planets \citep[for a similar strategy applied to transition disk systems, see][]{blakely2026}.

We begin in \S{2} with a discussion of the impact of a massive companion on a star's astrometry, illustrating with stars in \gaia~DR3 and the NASA Exoplanet Archive that have known companions (\S{3}). We focus on sources with distances, magnitudes, and color that are similar to stars in the \citetalias{pearce2022} catalog, and use both a measure of astrometric quality provided in \gaia~DR3 and a machine-learning algorithm based on astrometric errors and their correlations. In \S{4}, we apply our results to debris-disk systems and compare outcomes with the limits on planet mass and location provided by \citetalias{pearce2022}. We discuss our results in \S\ref{sec:discuss} and conclude in \S\ref{sec:conclude}.

\section{Astrometric signatures of companions in Gaia DR3}\label{sec:bg}

The astrometry of a star can reveal the presence of an unseen companion due to the star's reflex motion from the companion's gravitational force \citep[e.g.,][]{lattanzi2000, casertano2008}. In this section, we consider the general impact of orbital motion on astrometric measurements, its detectability, and how that motion might be detectable through measures of the quality of astrometric fits to linear drift across the sky.

A star moving on an otherwise straight path in the sky plane will acquire a periodic wobble that may impact the quality of a star's astrometry if the companion's influence is not considered. Estimates of position, parallax, and proper motion all fare poorly if a sequence of astrometric measurements of a star is fit with a model of linear drift, depending on the precision of the measurements and their cadence. 

This effect is well known. The \gaia\ DR3 archive includes a parameter \ruwe, the renormalized unit weighted error of a constant-drift astrometric model, which turns out to be a robust marker of binary and multiple stars \citep[e.g.,][]{brandt2018, brandt2021, belokurov2020, fabricius2021, gaiabin2023, kervella2022}. Stars with \ruwe\ above a threshold value of $\sim$1.4 are good candidates as members of binary or multiple-star systems, provided the binary orbit and the timing of \gaia's snapshots of the system are favorable. \citet{fitton2022} argue that the \ruwe\ threshold for binarity around stars with circumstellar material could be higher ($\sim$2.5), because they identified single stars with disks that have a higher \ruwe\ than those without circumstellar material. (Our premise is that massive planets might explain this correlation.) \citet{castro-ginard2024} provide a more detailed description of the connection between \ruwe\ and \gaia's astrometric measurements. 

To estimate how a companion impacts astrometry, we compare the physical motion of a star with astrometric sensitivity. When astrometric data are snapshots of sky positions --- ignoring radial velocity --- orbital motion affects astrometry only when it repositions a star by an amount larger than the positional uncertainty during the time interval over which multiple observations are made, $\Delta t$ \citep[cf.][]{lattanzi2000}. We consider two limiting cases.  First, at small separation when $\Delta t$ is large enough to cover a full orbit, detectability requires an orbital distance that is characterized by a semimajor axis of
\begin{eqnarray}\label{eq:alower}
    a & \gtrsim &  \frac{(M+\Mcompanion)}{\Mcompanion} \frac{\delta\varpi}{\varpi} \\ \nonumber
     \ & \gtrsim & 0.004 \times \frac{(1+q)}{q} 
    \left[\frac{\varpi/\delta\varpi}{250}\right]^{-1} 
    \text{\au},
\end{eqnarray}
where $M$ is the mass of the observed partner, which we assume to be the primary, $\Mcompanion$ is the companion/secondary mass, $\varpi$ is the true parallax (as in the \gaia\ archive we use units of mas), and $\delta \varpi$ is the typical error for sources at that parallax. In the lower expression, the mass ratio $q = \Mcompanion/M$ and the fiducial value of parallax-over-error is typical of the sources we consider below. This expression highlights that for planetary masses with $q \lesssim 10^{-2}$, orbital separation for nearby stars must exceed $a \sim $0.4~\au.  

Along with spatial resolution, the duration of observations relative to the length of the orbital period, $T_\text{orb}$, is key to astrometric sensitivity. Equation~(\ref{eq:alower}) holds when $\Delta t\gtrsim T_\text{orb}$; for an equal-mass Sun-like binary with $a \gtrsim 0.01$~\au, observations must span at least $T_\text{orb} \gtrsim 1~\text{d} \sim 10^{-3}$~yr.  Below, we often discuss astrometric sensitivity in terms of orbital period, with less concern for spatial separation between a star and its companion.

Measurements are complicated at small separation when stellar partners are not resolved and only the center of light is observed. With identical twin partners, the position of the center of light does not change at all. When one companion is faint or dark \cite[e.g.,][for black holes]{muller-horn2025}, the orbital motion of the brighter star may be detectable. However, rapid orbital motion may well appear as random noise that does not strongly impact an astrometric fit to a linear drift model. For stars in \gaia\ that we consider here, there are sets of several observations made over the course of a few days, and dozens of such sets (as indicated by Gaia's \vpu\ parameter) spanning several years. We conjecture that even with time-series data, orbital motion more rapid than this cadence may be buried in other noise sources. We thus adopt a lower limit of 0.1~yr for the orbital period of a stellar binary to impact linear-drift astrometry. As in the next section, many binary sources with periods below that value give reasonable fits to a constant-drift astrometric model ($\ruwe < 1.4$).

The requirements for astrometric impact for a low-mass companion are even more stringent. From Equation~(\ref{eq:alower}), 
\begin{equation}\label{eq:planetperiodlower}
    T_{orb} \gtrsim 2.8{\times}\left[\frac{q}{0.002}\right]^{\!-3/2} 
    \left[\frac{\delta \varpi/\varpi}{0.004}\right]^{3/2}
    \left[\frac{M}{{1~\msun}}\right]^{\!-1/2}\!\text{yr}. 
\end{equation}
Shorter-period planets do not impact the observed motion of their host star as the star's positional changes about the center of mass are too slight to be detected.

For long-period orbits, binary motion may be detected as an acceleration, depending on the resolution of proper motion $\vec{\mu}$ \citep[e.g.,][]{brandt2018, brandt2021}.  By assuming that the orbital acceleration is nearly constant over the observation time $\Delta t$, the velocity deflection $\Delta v$ in this time frame is
\begin{equation}
    \frac{\Delta v}{\Delta t} \sim \frac{G \Mcompanion}{a^2} ~ ,
\end{equation}
where $G$ is Newton's gravitational constant. To translate $\Delta v$ into a deflection in proper motion, $\Delta \mu$, we connect physical speed with the proper motion error, $\delta\mu$. Then, a linear drift model could discern a velocity deflection if $\Delta v \approx 2 \delta \mu$, so long as $\Delta\mu$ stands out above expected errors in the fitting. From this estimate, we find an upper bound to the orbital separation,
\begin{eqnarray}
a & \lesssim &  \left[\frac{G q \Mcompanion\,\Delta t}{D\, \delta\mu}\right]^{1/2}
\\  \nonumber
\ & \lesssim & 
120 \, q^{1/2} \left[\frac{\varpi}{5~\text{mas}}\right]^{1/2}
    \left[\frac{\delta\mu}{0.04~\text{mas/yr}}\right]^{-1/2} \\  \nonumber
    \ & \ & \ \ \ \times \ 
    \left[\frac{M}{1~\msun}\right]^{1/2} 
    \left[\frac{\Delta t}{3~\text{yr}}\right]^{1/2}
    \text{\au,}
\end{eqnarray}
where $D = 1/\varpi$ is the source's heliocentric distance in units of kiloparsecs when the units of parallax $\varpi$ are milliarcseconds. Converting this value to the companion's orbital period, we estimate that for nearby equal-mass, Sun-like stars, binary motion impacts \gaia\ astrometry when the orbital period is less than about 950~years. When the binary mass ratio drops to $q=0.1$, the bound is about 200~years. 

For planetary companions, the upper limit to the orbital period for astrometric impact is roughly
\begin{eqnarray}\label{eq:planmetperiodupper}
    T_{orb} & \lesssim & 13 \, \times 
    \left[\frac{q}{0.002}\right]^{3/4} 
    \left[\frac{\delta\mu}{0.04~\text{mas/yr}}\right]^{-3/4}
    \\ \ & \ & \ \ \  \nonumber
    \left[\frac{\varpi}{5~\text{mas}}\right]^{3/4}
    \left[\frac{M}{1~\msun}\right]^{1/4} 
    \left[\frac{\Delta t}{3~\text{yr}}\right]^{3/4} \ \text{yr}.
\end{eqnarray}
A Sun-like star with a companion of mass $\Mcompanion = 2\mjup$ (where \mjup\ is the mass of Jupiter) gives an upper bound close to the lower bound in Equation~(\ref{eq:planetperiodlower}). The implication is that for nearby Sun-like stars measured by \gaia, astrometric quality indicators like \ruwe\ are sensitive to planetary companions if the planets are roughly as massive as Jupiter and located at orbital distances comparable to those of the Sun's gas giants. 

In formulating the results of this section, we were guided by \gaia\ binary star observations. Extending an astrometric analysis to potential planetary hosts is a challenge because of the low mass ratios. Among the 178 debris-disk systems from \citetalias{pearce2022}, only about 20\%\ have predicted planets with a minimum mass of 1~\mjup\ or more. The predictions are just lower limits; if all 178 stars in the sample have masses that are three times the predicted value \citep[Fig.~7 therein]{pearce2022}, then more than half of the systems contain a planet that is at least as massive as Jupiter. First, to set the stage for this analysis, we explore binary star data.

\section{Astrometric measurement of stars with companions}\label{sec:knowncomps}

In this section, we examine stars similar to the debris-disk systems of \citetalias{pearce2022} to assess how \gaia\ DR3 astrometry, in particular \ruwe, performs as an indicator of companions. We first establish selection criteria to obtain samples of stars with observable properties similar to those in the \citetalias{pearce2022} catalog. We review the connection between binarity and \ruwe, showing how the metric responds for stars with these selection criteria and validating our analysis in \S\ref{sec:bg}. We then focus on well-studied planet hosts that also meet these criteria to establish how \ruwe\ is distributed in the absence of any stellar or massive planetary companion. Finally, with the set of known planet hosts, we look for evidence that \ruwe\ and other astrometric indicators might reveal planetary companions. We begin by establishing selection criteria that characterize the observed properties of the debris-disk systems.

The quality of astrometric fits can be affected by the observed characteristics of individual sources, including parallax, proper motion, brightness, and color. Therefore, we first consider these characteristics for stars that are of ultimate interest here, the sources in the \citetalias{pearce2022} catalog that were observed by \gaia. We obtain \gaia\ DR3 identifiers of sources in the \citetalias{pearce2022} catalog from SIMBAD \citep{wenger2000}. All but two of the 178 stars are in the \gaia\ DR3 data set. The two stars missing in the archive are the A-type star $\beta$~Leo and the B-type star $\epsilon$~Sgr. Both are too bright for \gaia\ astrometry, with visual magnitude V $\approx$ 2, as reported in SIMBAD. The remaining stars constitute the 176 debris disk hosts that we study here. 

Our next step is to identify other stars --- both with and without stellar or planetary companions --- that have similar observed characteristics. To do so, we estimate selection criteria from the 176 debris disk systems, all of which satisfy these conditions:
\begin{gather}\label{eq:sel}
2~\text{mag} \leq \photgmeanmag \leq 12~\text{mag}, \\  \nonumber
-0.25~\text{mag} \leq \bprp < 3.5~\text{mag},  \\  \nonumber
\ \ \parallax > 4~\text{mas}, \ \ 
\text{and} \\  \nonumber
- 1.5~\text{mag} \leq M_G - 3.5 \times \bprp < 3.5 ~\text{mag}, 
\end{gather}
where \photgmeanmag\ is the \gaia\ G-band apparent magnitude, \bprp\ is the color in \gaia\ blue and red bands, \parallax\ is the \gaia\ parallax from the main source catalog, and $M_G$ is the G-band absolute magnitude derived from the apparent magnitude and parallax \citep{gaiadr3}. Our choices for the parameters in Equation~(\ref{eq:sel}) are designed to broadly admit stars that are of similar brightness and parallax as the debris-disk systems, since both attributes can affect astrometry. We also select by color and absolute magnitude to identify stars on or close to the main sequence, similar to the 178 systems. These criteria select A-type to M-type main-sequence stars that are nearby (within 250~pc) and optically bright (less than 12$^\text{th}$ magnitude in \gaia\ G-band).

We apply the criteria in Equation~(\ref{eq:sel}) to a well-studied set of planet hosts in the NASA Exoplanet Archive \citep[``NXA'' hereafter]{nxa2013}, each of which hosts a confirmed exoplanet. A subset of 1,117 of these stars meet the selection criteria in Equation~(\ref{eq:sel}); these stars all have observational characteristics similar to the \citetalias{pearce2022} sources and will be part of our planetary companion analysis below.

\begin{figure}
    \centering
    \includegraphics[width=1\linewidth]{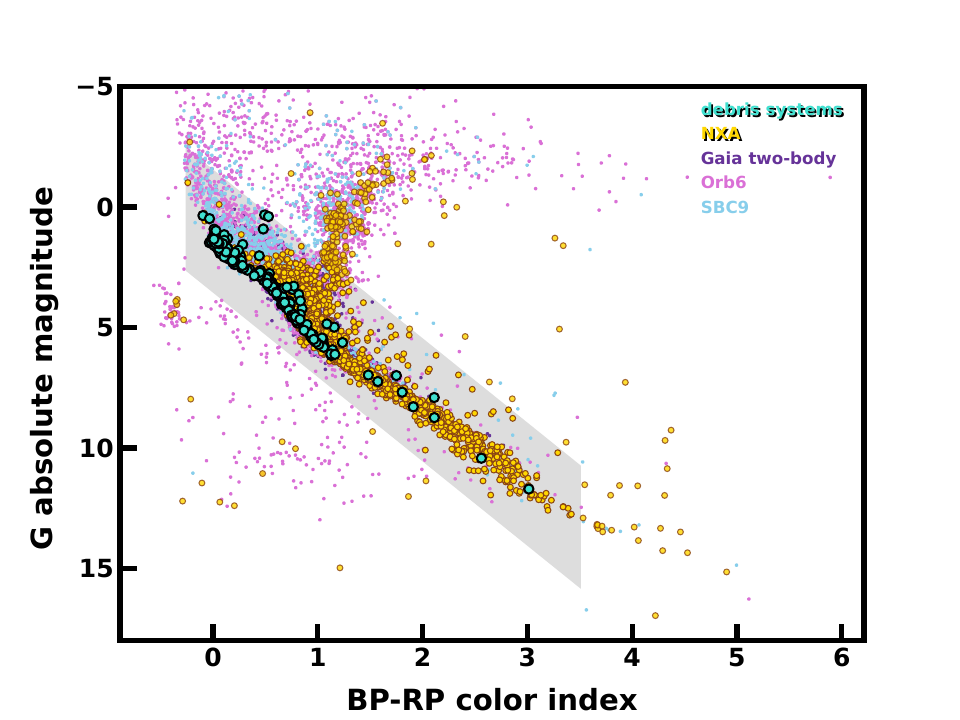}
    \caption{Hertzsprung-Russell/color-magnitude diagram for 176 debris disk host stars (turquoise) and confirmed exoplanet hosts (gold; ``NXA''). The shaded region defines a ``selection zone'' so that when we compare collections of stars we only consider those that are predominantly on the main sequence, and with a range of color and magnitude similar to the set of 176 debris disk systems. The plot includes binary sources from the Gaia two-body table (purple), and stars from the SBC9 (blue) and Orb6 (pink) catalogs. In our analyses, we only use sources that lie in the selection zone.}
    \label{fig:cmd}
\end{figure}

\begin{figure}
    \centering
    \includegraphics[width=1\linewidth]{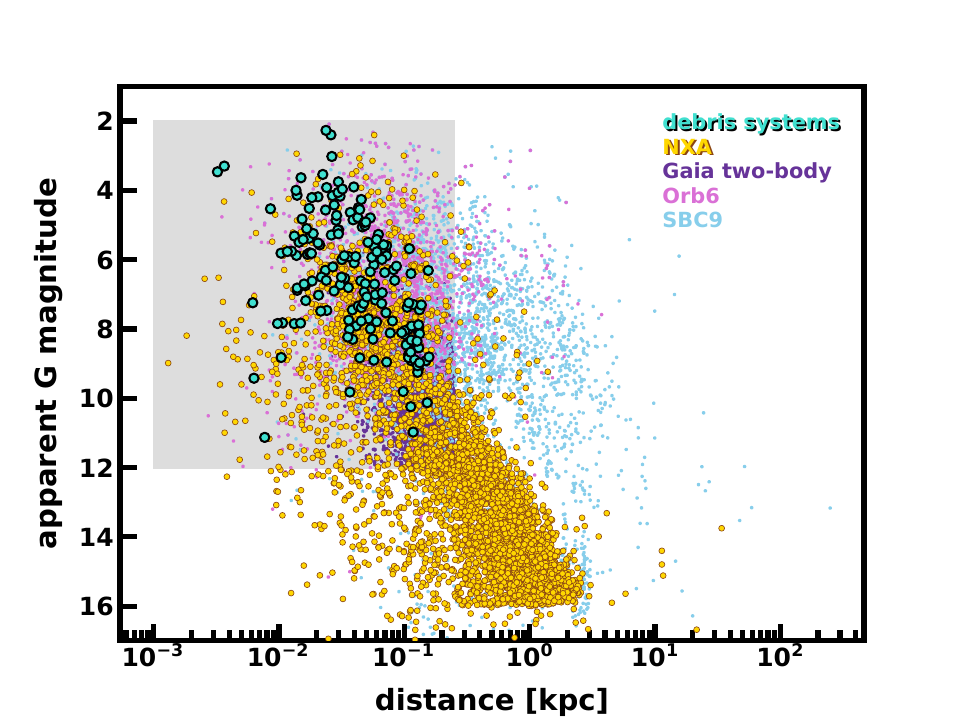}
    \caption{Distance and apparent magnitude of debris disk host stars (turquoise) and exoplanet hosts from the NASA Exoplanet Archive (gold). Distances for these nearby sources are determined directly from the inverse of the parallax. A comparison with \citet{bailerjones2021} distance estimates shows that when the latter are available (approximately half of the sources), the difference is small, well within a percent for these nearby sources. Also shown are Gaia two-body sources (purple) and stars from the SBC9 and Orb6 catalogs, as in Figure~\ref{fig:cmd}.}
    \label{fig:paxmag}
\end{figure}

Figures~\ref{fig:cmd} and \ref{fig:paxmag} show observational characteristics of the 176 debris-disk stars, as well as the NXA planet hosts. The first figure is a Hertzsprung-Russell diagram; the second shows parallax versus apparent magnitude. A gray region in each figure shows where the constraints in Equation~(\ref{eq:sel}) are satisfied. They constitute a ``selection zone'' that encompasses all of the 176 stars with debris disks, and includes the 1,117 NXA stars.

Figures~\ref{fig:cmd} and \ref{fig:paxmag} also show three sets of sources from binary star catalogs, included in our analysis to illustrate the general connection between \ruwe\ and binarity for stars with similar brightness, color, and distance as the debris disk systems (Eq.~(\ref{eq:sel})). The first set is 1,492 binary stars from the U.S.~Naval Observatory Double Star catalog, with derived orbital elements \citep[Orb6]{ORB62001a, ORB62001b}. The second set consists of 1,617 sources from the Ninth Catalog of Spectroscopic Binaries \citep[SBC9]{SBC2004}. The third contains 3,678 randomly selected stars in \gaia\ DR3 that have $\nss = 0$. These random sources represent single stars, though they are better labeled ``not-non-single stars,'' in acknowledgment that \gaia's non-single-star tables have strict admission standards that may miss stellar partners as well as planetary companions \citep{gaiamulti2023}. The stars identified in all three catalogs are in both SIMBAD and the \gaia\ DR3 archive, and all meet the criteria in Equation~(\ref{eq:sel}). They were further culled to limit their proper motion in \gaia\ DR3 to within 1,000~mas/yr, so that astrometric quality might be less affected by large changes in position over the course of \gaia's observations.

\begin{figure}
    \centering
    \includegraphics[width=1\linewidth]{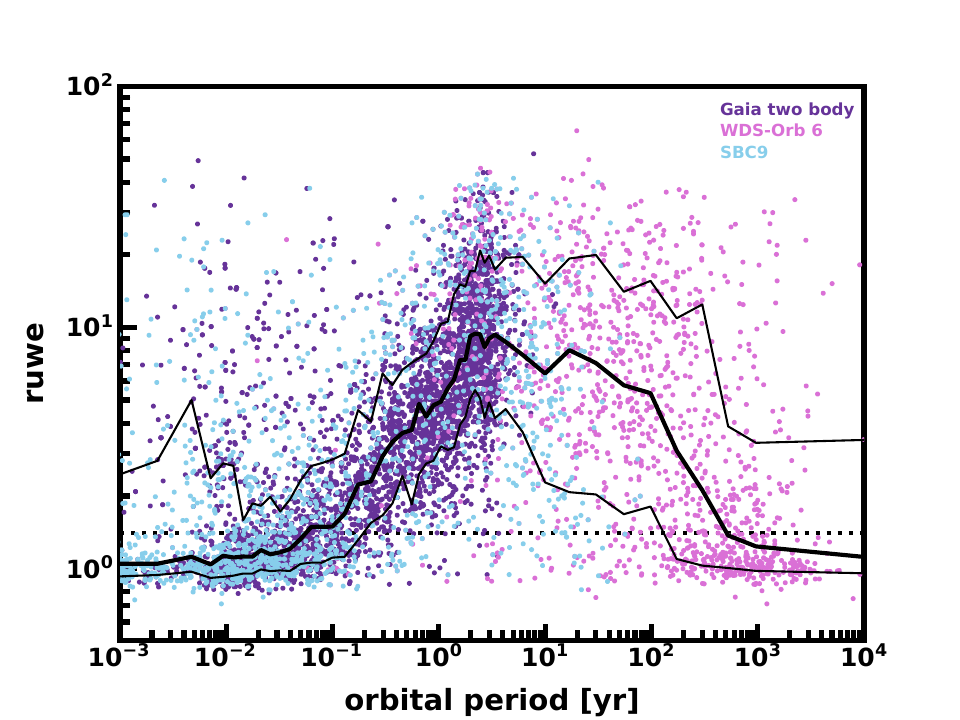}
    \caption{\gaia's astrometric quality measure, \ruwe, as a function of binary orbital period. In a ``sweet spot'' of orbital periods ranging from 100 days to 100 years, \ruwe\ is strongly sensitive to binary motion. Outside of the sweet spot, the binaries are unresolved or vary only by small amounts on the plane of the sky compared with the (linearly drifting) binary center of mass. The solid lines indicate the 16th, 50th and 84th percentiles in populations binned by orbital period.}
    \label{fig:ruwe}
\end{figure}

Toward connecting binarity and \ruwe, we illustrate the impact of stellar companions on astrometry in Figure~\ref{fig:ruwe}, which shows the quality metric \ruwe\ as a function of the companion's orbital period. For reference, $\ruwe \lesssim 1.4$ indicates a good fit to the model of a single star drifting with constant proper motion on a line across the sky \citep{gaiabin2023}. The effect of the observational time baseline ($\Delta t$ in \S\ref{sec:bg}) on the sensitivity of \ruwe\ to binary motion is clear. Binaries with orbital periods less than about a month are so close that their overall motion in the sky is fairly well-fit by a linear drift model about the center of mass. Stars with companions that have orbital periods exceeding a century may also appear to be linearly drifting, since their orbital motion is small over the 3-yr duration of \gaia\ observations.  In between, in a sweet spot of orbital periods,
\begin{equation}\label{eq:sweet}
    0.1~\text{yr} \leq T_\text{orb} < 100~\text{yr},
\end{equation}
binary motion causes a significantly bad fit to the linear drift model. We hereafter adopt the term ``sweet spot'' to designate this range of orbital periods.

In quantitative terms, stars in Figure~\ref{fig:ruwe} with stellar companions on orbits with periods between 0.1 and 100 years have a median \ruwe\ value of 5.4. About 94\%\ of this population have $\ruwe > 1.4$, near the threshold between a good and poor fit to linear drift. For stars in binaries with shorter periods, the fraction of objects with $\ruwe > 1.4$ drops to around 34\%, while 54\%\ of stars in longer-period binaries have \ruwe\ above that threshold. Evidently, binarity may well impact the quality of a linear-drift astronomical solution even at short and long orbital periods \citep[cf.][]{brandt2021, castro-ginard2024, elbadry2025}. However, if we wish to use \gaia\ DR3 astrometry to detect or confirm binary motion, our sensitivity will be greatest for sources in the sweet spot with $0.1 < T_\text{orb}/\text{yr} < 100$. These conclusions are consistent with the assessments and discussion in the previous section (\S\ref{sec:bg}).

To further strengthen the connection between \ruwe\ and binary motion, we identify a population of planet hosts in the NXA (exoplanet archive) that meet our selection criteria (Eq.~(\ref{eq:sel})), obey the proper motion ``speed limit'' of 1,000~mas/yr (as above), and do not have SIMBAD indicators of stellar partners. Specifically, we admit only sources for which the NXA's listed number of stars in a host's planetary system is $\sysnum = 1$. We also select stars with \gaia\ DR3's $\nss = 0$. This flag has a high bar for admission, and is generally an indicator of a binary partner, a substellar companion, or even a massive giant planet \citep{stefansson2025}. Although we may miss potential discoveries by excluding sources with measured acceleration, our goal here is to focus on planetary systems that are not yet resolved by \gaia.

Furthermore, we require that SIMBAD's object-type list \otypes\ does not contain a double star flag (\texttt{**}, \texttt{SB*}, or \texttt{EB*}).\footnote{A check on \otypes\ stored in a Python Pandas dataframe could have the form \texttt{df.otypes.str.contains('**', regex=False)}.} Lastly, we take the maximum mass ratio between planet and star to be less than 0.3~\mjup/\msun\ to avoid the impact of a known giant planet on the star's astrometry (\S\ref{sec:bg}). Our final tally of these companionless NXA stars (hereafter ``solo stars'') is 388 sources.

The median \ruwe\ value of solo NXA stars is within a percent of unity, and the fraction of stars with $\ruwe > 1.4$ is just under 3\%. The $95^\text{th}$ percentile in \ruwe\ is 1.3, and the maximum value is 1.94,  corresponding to WASP-131, a $\sim$10~mag G0 star  at about 200~pc from us \citep{nxa2013}. It hosts a 0.27~\mjup\ planet (at the high end of the planetary masses for the solo stars) orbiting at about 12~\rsun (equivalent to 0.3~mas).   

We summarize the statistics of these stars and of known binaries in Table~\ref{tab:ruwestats}.

\begin{deluxetable}{crrrrrc}
\tabletypesize{\footnotesize}
\tablecaption{Statistics of \ruwe\ in binary stars
\label{tab:ruwestats}}
\tablehead{
   \colhead{} & 
   \colhead{} & 
   \multicolumn{3}{c}{\ruwe\ percentile} & 
   \colhead{\ruwe} &
   \colhead{\ruwe$>$1.4} 
\\
  \colhead{Sample} & 
  \colhead{N} &
  \colhead{5$^\text{th}$} & 
  \colhead{50$^\text{th}$} &
  \colhead{95$^\text{th}$} & 
  \colhead{max} & 
  \colhead{(pct.)}}
\startdata 
short-period & 1894 & 0.89 & 1.16 & 7.21 & 49.19 & 34.0 \\ 
long-period & 669 & 0.93 & 1.55 & 18.81 & 37.31 & 54.1 \\ 
sweet-spot & 3417 & 1.27 & 5.39 & 23.49 & 65.45 & 93.6 \\ 
not-nss & 3679 & 0.82 & 1.02 & 4.32 & 49.30 & 16.4 \\ 
solo NXA & 388 & 0.84 & 1.00 & 1.30 & 1.94 &  2.8  
\enddata
\tablecomments{Short-period binaries have orbital periods of less than 0.1~yr, long-period binaries have periods exceeding 100~yr, and ``sweet-spot'' binaries have orbital periods in between. The ``not-nss'' stars are the random sample from \gaia\ with $\nss= 0$, as described in the text. The solo NXA stars are selected to have no stellar or giant planetary companions.}
\end{deluxetable}

The importance of working with sources like the 391 solo stars from the NXA that have been well-studied in terms of companions is underscored when we consider the ``not-non-single'' stars with $\nss = 0$. In our random download from \gaia\ DR3, this set has a median \ruwe\ value of 1.02, and over 16\% of the sources have $\ruwe > 1.4$. The $99^\text{th}$ percentile and maximum \ruwe\ values are 16.6 and 49.5, respectively. We interpret these results to mean that \gaia's $\nss$ flag is intended to be selective, not all-encompassing \citep{gaiadr3}, so that our subsample is expected to contain some binaries. 

\begin{figure}
    \centering
    \includegraphics[width=1\linewidth]{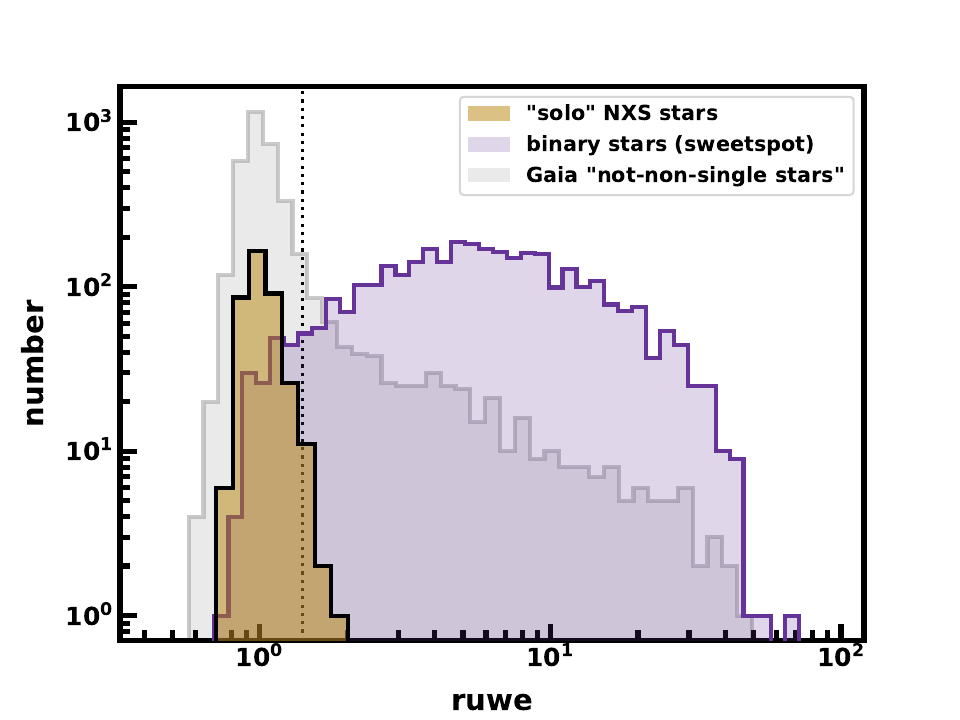}
    \caption{Histograms of \gaia's astrometric quality measure, \ruwe, for several star populations. The black/gold histogram indicates the \ruwe\  distribution for NXA single stars, with no evidence of a stellar or a gas giant companion. The violet/pink histogram corresponds to known binaries with orbital periods in the sweet spot. The power of \ruwe\ as a discriminator of binary motion is evident in these distributions. The value of 1.4 (black dotted line) is the threshold value commonly adopted in the literature. The gray histogram is from \gaia\ stars with $\nss = 1$, and the significant fraction (16\%) of these sources with $\ruwe > 1.4$ is, we conjecture, the result of binaries in this population.}
    \label{fig:ruwehist}
\end{figure}

These results are summarized in Figure~\ref{fig:ruwehist}, showing \ruwe\ distributions for the NXA single stars, the \gaia\ non-single stars in the orbital-period sweet spot, and the ``not-non-single'' stars.

\subsection{Comparison with predictions of astrometric quality}

As previous work demonstrates, $\ruwe > 1.4$ is a strong indicator of binary motion \citep{stassun2021, penoyre2022, gaiabin2023, castro-ginard2024}. Our own assessment of the orbital configurations of stellar binaries to which \ruwe\ is sensitive implicates a range from $10^{-3}$--$10^{3}$ years, with the caveat that the lower bound would need to be adjusted upward so that \gaia~astrometric sampling could track coherent orbital motion, suggesting roughly 100~days. Figure~\ref{fig:ruwe} confirms that this lower bound is indeed appropriate, in general, even though approximately half of binaries with shorter orbital periods have \ruwe\ values above 1.4.

Our upper limit for orbital periods of binaries that may impact astrometry, about 1,000 years, seems validated in Figure~\ref{fig:ruwe}. Just below $T_\text{orb} = 100$~yr, \ruwe\ values for binary stars are overwhelmingly above 1.4; for binaries with periods above about $10^4$~yr, the distribution is opposite. Our value of $T_\text{orb} = 100$~yr for the large-period limit of the sweet spot is a conservative estimate.

\subsection{Planetary companions}

\begin{figure}
    \centering
    \includegraphics[width=1\linewidth]{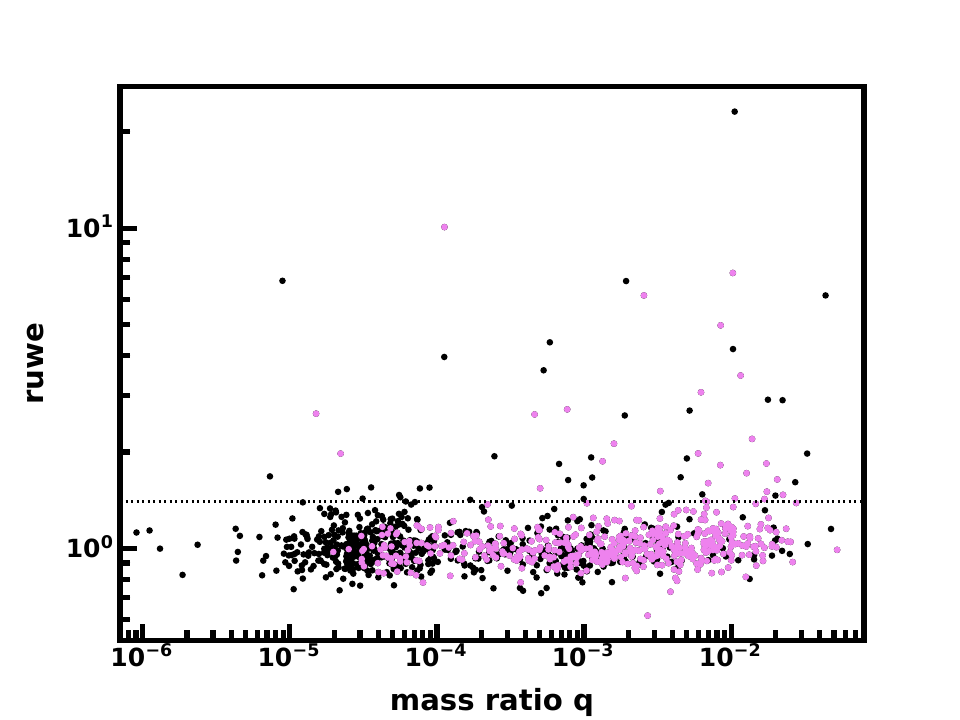}
    \caption{The astrometric quality parameter \ruwe\ versus companion mass ratio $q$ for NXA planet hosts. The 1,117 sources shown all match our selection criteria (Figures~\ref{fig:cmd} and \ref{fig:paxmag}). The value of $q$ corresponds to the most massive planet if a host has multiple planetary companions. The lighter colored points have planets with orbital periods in the ``sweet spot''.}
    \label{fig:nxa_qpl_ruwe}
\end{figure}

The NASA Exoplanet Archive contains over 6,000 confirmed planets around over 4,200 host stars, including 1,117 sources that match our selection criteria illustrated in Figures~\ref{fig:cmd} and \ref{fig:paxmag}. This resource offers an opportunity to explore the sensitivity of astrometric quality measures to the presence of a planet. Figure~\ref{fig:nxa_qpl_ruwe}, showing \ruwe\ as a function of the planet-to-star mass ratio $q$ for each host's most massive planet offers a preview.

As above, we consider our 388 solo NXA sources that show no sign of having a binary stellar or massive planetary companion. We also look at sets of 367 ``single'' NXA stars that include a massive planetary companion but no stellar companion to identify trends in \ruwe\ statistics. Table~\ref{tab:plprobs} highlights our results. When we divide the NXA stars into these two groups, solo stars versus single stars with massive planets (mass ratio $q > q_\text{cut} = 0.3$ in units of \mjup/\msun), the \ruwe\ distributions are similar. We cannot rule out that the populations are different on the basis of a two-sample Kolmogorov-Smirnov (KS) test. A test of the proportion of stars with $\ruwe > 1.4$, the proportion-$z$ (``prop-$z$'') test, also fails to detect a statistically significant difference between these two samples.
We conclude that planetary hosts with these demographics do not reveal themselves through \ruwe\ statistics.

\begin{deluxetable}{ccccc}
\tabletypesize{\footnotesize}
\tablecaption{Tests of \ruwe\ distributions for planet hosts with a range of planetary masses.
\label{tab:plprobs}}
\tablehead{
  \colhead{mass ratio} & 
  \colhead{star} &
  \colhead{\ruwe$>$1.4} & 
  \colhead{KS} &
  \colhead{prop-$z$} \\
  \colhead{ $q_\text{cut}$ \scriptsize{(\mjup/\msun)}} & 
  \colhead{count} & 
  \colhead{count} & 
  \colhead{$p$-value} & 
  \colhead{$p$-value } }
\startdata 
  0.3 &  367 & 17 & 0.2685 & 0.192 \\
    1 &  253 & 14 & 0.0660 & 0.085 \\
    2 &  180 & 11 & 0.0286 & 0.060 \\
    5 &   93 & 10 & 0.0071 & 0.001 \\
   10 &   43 & 7 & 0.0021 & $< 0.001$ 
\enddata
\tablecomments{The first column ($q_\text{cut}$) indicates the lower bound to a range of planet-to-star mass ratios ($q$). The second column is the number count of stars whose most massive planetary companion falls in that $q$ range. The third and fourth columns highlight the differences in the \ruwe\ distribution of each set of stars as compared with NXA single stars with no giant planets. The third column contains KS test $p$-values that measure the differences in the two \ruwe\ distributions, while the fourth column is a prop-$z$ test assessing whether \ruwe\ from the set of stars is different from the NXA singles with no giant planets on the basis of the number counts with $\ruwe > 1.4$.}
\end{deluxetable}

\begin{figure}
    \centering
    \includegraphics[width=1\linewidth]{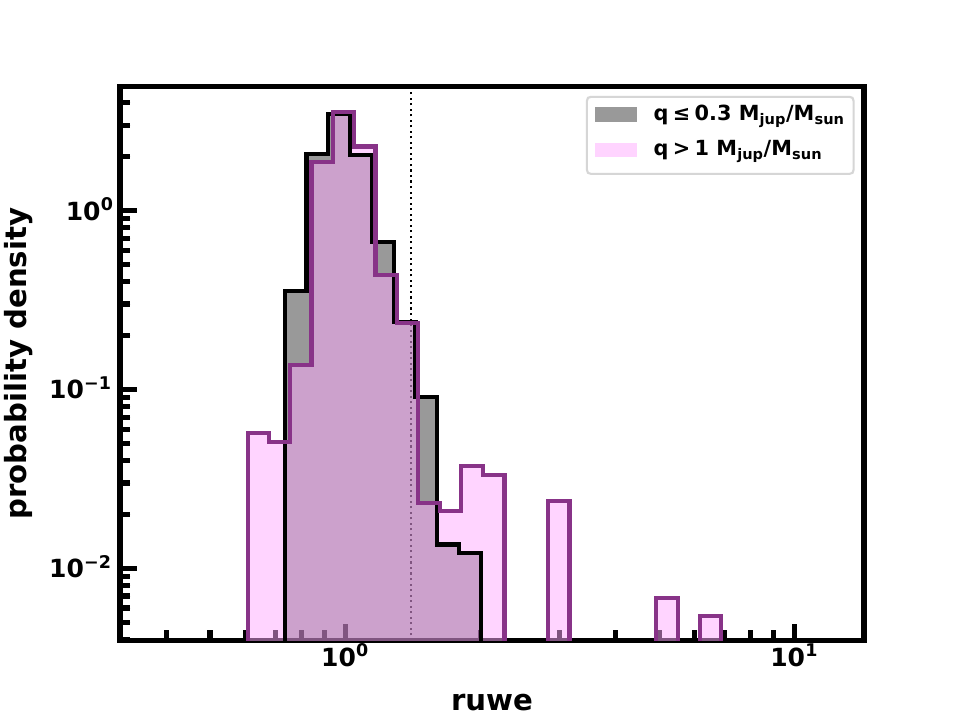}
    \caption{Histogram of \ruwe\ for single host stars with and without a massive planet. The stars with planet mass ratios $q \ge 0.3 \mjup/\msun$ are indicated by the purple histogram, while stars with lower-mass planets are indicated in the gray histogram. The tail of the histogram for $q \ge 0.3 \mjup/\msun$ is notable, serving as an indicator that \ruwe\ may be sensitive to planetary companions.}
    \label{fig:ruweconfirmedplanetshist}
\end{figure}

The situation changes as we increase the minimum mass ratio $q_\text{cut}$. When the threshold value is equivalent to 2~\mjup\ about a solar-mass star, the KS test suggests a significant difference between the \ruwe\ distribution for single stars with a massive planet compared to solo stars ($p\text{-value}= 0.029$). The prop-$z$ test gives a marginal $p$-value of about 6\%. For a threshold equivalent to 5~\mjup, both tests show a significant difference between the two populations. At the highest value of $q_\text{cut} = 10$~\mjup/\msun, the KS test $p$-value is about 0.2\%, while the prop-$z$ test has a $p$-value less than $10^{-4}$. On the basis of the 43 stars with massive planets in this range (Fig.~\ref{fig:ruweconfirmedplanetshist}), we conclude that astrometric quality may be a potential signature of planet host candidates.

To check that the \ruwe\ excess among hosts of massive planets might stem from apparent magnitude,  parallax or proper motion selection, we binned according to objects above or below the median values in these quantities. The \ruwe\ distributions of these subsets show variations. Closer sources tend to have more \ruwe\ outliers compared to the NXA single stars with no giant planets, while the proper motion and brightness selection do not affect the \ruwe\ distributions as much.  We caution that the number of sources is small in some of these subsets, particularly for hosts with massive planets.

\subsection{Beyond RUWE: a machine learning approach}\label{sec:ml}

While \ruwe\ provides a useful summary diagnostic of excess astrometric scatter, the \gaia\ DR3 archive has additional information that may help to identify unseen bound companions. For example, orbital motion may increase uncertainties in parallax and proper motion, possibly introducing tell-tale correlations in those errors that are not perceptible with \ruwe. We therefore consider a suite of parameters including parallax and proper motions, the errors in each parameter, and their mutual error correlations (e.g., \corrcoeff{pmra}{pmdec}). We also add to the analysis error correlations between proper motion components and sky position along with \gaia\ brightness (\photgmeanmag) and color (\bprp).

To explore whether this larger list of features can reveal companions to stars in \gaia, we adopt a machine-learning (ML) approach. We begin with a training set of 755 stars in the NXA for which there is no evidence of stellar binarity. About a third of this group has massive planets in the ``sweet spot'' based on the reported period $T_\text{orb}$ and mass ratio $q \ge 0.3$ (in units of $\mjup/M$ where $M$ is the host mass). The remaining objects are ``solo,'' with lower-mass planets only. Using random draws (Pandas' \texttt{sample} dataframe method) we up-sample the smaller population to have equal numbers of both types, with label $\ell = 0$ for solo, and $\ell = 1$ for hosts of larger-mass planets. This procedure yields a set of 1,008 elements.

From this training set, we gather features from parameters in \gaia's main source archive. Our complete list is in Table~\ref{tab:gaiaparms}. Our approach thus includes information that may be hidden in \ruwe, for example, if orbital motion (as opposed to random errors) induces correlations in errors of parallax and proper motion, depending on how a source moves across the sky. We also include photometric data, which may impact how planetary companions impact astrometry. For example, with our selection zone that isolates the main sequence, color correlates with host mass, and therefore may be an indicator of potential reflex motion.

\begin{deluxetable*}{llc}
\tabletypesize{\footnotesize}
\tablecaption{\gaia DR3 parameters in the machine-learning analysis
\label{tab:gaiaparms}}
\tablehead{
  \colhead{parameter} & 
  \colhead{description} & 
  \colhead{importance}}
\startdata 
\parallax & parallax [units: mas] & 0.1134 \\
\paxerr  & [1-$\sigma$ errors; mas] & 0.0566 \\
\pmra       &   proper motion, along right ascension [mas/yr] & 0.0498 \\
\pmraerr   & [mas/yr] & 0.0428 \\
\pmdec    & proper motion, along declination [mas/yr] & 0.0449 \\
\pmdecerr  & [mas/yr] & 0.0513 \\
\corrcoeff{ra}{parallax} & correlation coefficient [unitless]  & 0.0321 \\
\corrcoeff{ra}{pmra} &  \multirow{2}{*}{\ \ \ $\vdots$} & 0.0384 \\
\corrcoeff{ra}{pmdec} &  & 0.0328 \\
\corrcoeff{dec}{parallax} &  & 0.0306 \\
\corrcoeff{dec}{pmra} &  & 0.0332 \\
\corrcoeff{dec}{pmdec} &  & 0.0405 \\
\corrcoeff{parallax}{pmra} &  & 0.0357 \\
\corrcoeff{parallax}{pmdec} &  & 0.0379 \\
\corrcoeff{pmra}{pmdec} &  & 0.0353 \\
\ruwe         &   renormalized unit-weighted error [unitless] & 0.0402 \\
\photgmeanmag &  G-band apparent magnitude [mag] & 0.2104 \\
\bprp        &  blue-red color [mag] & 0.0740 
\enddata
\tablecomments{The list is in order of appearance in the \gaia\ source archive. }
\end{deluxetable*}

In our supervised learning approach, we use Python Scikit-Learn's random forest algorithm (\texttt{RandomForestClassifier} in the \texttt{sklearn} module, version 1.2.2). Our implementation adopts the classifier's default parameters, including 100~trees with no set maximum depth to the tree branches, based on experience with previous work \citep{whiting2023}. The goal of the classifier is to assign a probability $\probpl$ that a source is a planet host, and we further reduce the output to a predicted label $\ell_\text{pred}$ for each source, depending on that likelihood: $\ell_\text{pred}$ is one if $\probpl > 0.5$, and zero otherwise. We quantify accuracy, or success, as the fraction of sources that are correctly identified as planet hosts ($\ell_\text{pred} = \ell$).

We performed train-test trials with random partitions of the up-sampled training set, exploring subsets of the features listed above. The full list of features performed best, with train-test trials yielding about an 80\%\ success rate when applied to a test set without up-sampled elements, nominally better than random.

We also tested a neural net classifier (\texttt{sklearn}'s multi-layer perceptron algorithm \texttt{MLPClassifier} with default parameters and variations in the number of perceptron layers and the regularization factor $\alpha$) on the same features (in this case rescaled to unit variance with \texttt{transform} in the \texttt{preprocessing} module under \texttt{StandardScaler}) to identify if a neural net would perform better. It did not.  For example, the default multilayer perceptron parameters yielded around 77\%\ accuracy in test-train trials.

To generate our final ML model, we train the random forest classifier on the full set of 1,008 sources, including the up-sampled massive-planet hosts. This model is the basis of our analysis of the \citetalias{pearce2022} catalog. 

With machine learning approaches, it is sometimes unclear how a classifier is operating. This conclusion is generally true of our methods; however, we gain insight by exploring the random forest classifier's ``feature importances,'' which indicate how much weight each feature has in the classification scheme. In this study, the classifier weights brightness, parallax, and color most heavily (39.8\%), then proper motion, \ruwe, and other astrometry errors (28.7\%), and finally the astrometric correlation coefficients (31.6\%). Photometric features, brightness and color, account for 28.4\% of the importance weighting, while astrometric features account for the rest (71.6\%). The non-astrometric features give the classifier access to observed and gross stellar properties that are ultimately connected to stellar mass, information that may be valuable in interpreting the astrometric data. A liability of including non-astrometric features like brightness and color is that they may imprint the classifier with the selection biases in planet detection techniques. Still, the broader range of features is important to explore, as only a minority of actual planet hosts have high \ruwe\ values.

\section{178 debris-disk systems (minus two)}\label{sec:176}

\begin{figure}
    \centering
    \includegraphics[width=1\linewidth]{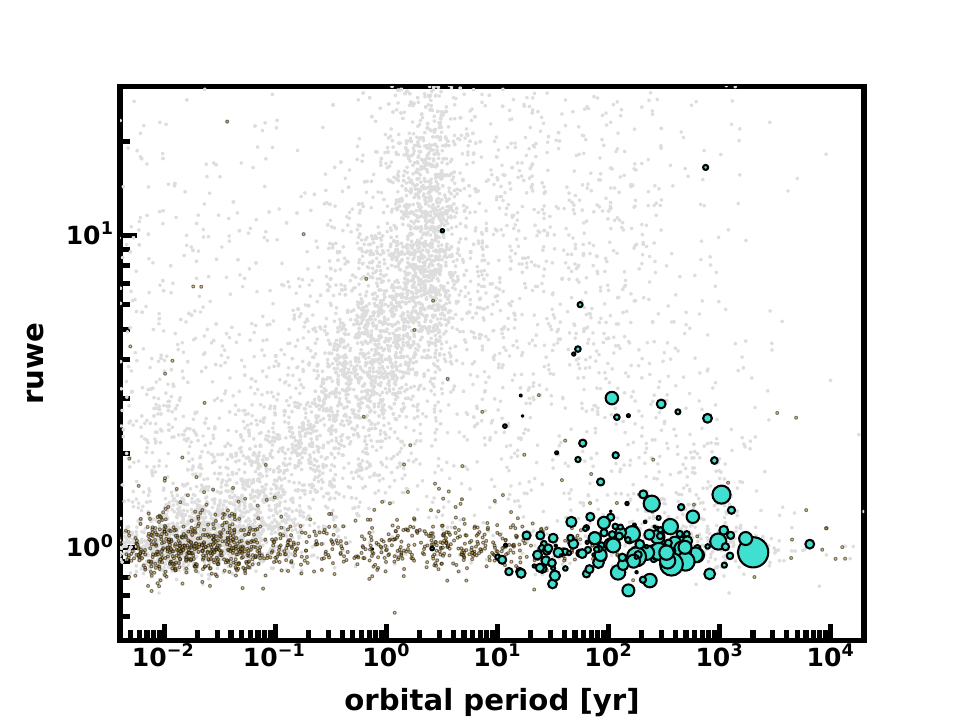}
    \caption{The maximum orbital period of potential planets and \ruwe\ of the 176 debris-disk systems from \citetalias{pearce2022}. The turquoise points correspond to these sources; the diameter of each symbol is proportional to the predicted minimum mass. For reference, points from the NXA planet hosts (gold) and binary stars (faint light gray) are also shown. The debris-disk systems' inferred upper bound of the planetary orbital periods lie in the ``sweet spot,'' although typically at higher values than the planet hosts.}
    \label{fig:176_ruwe_period}
\end{figure}

We now apply our analysis to the sample of 178 debris disk systems identified by \citetalias{pearce2022} to determine if there is astrometric evidence for planetary companions. \citetalias{pearce2022} predict the minimum mass $\mmin$ and maximum orbital distance (from which we derive a maximum period $T_\text{max}$) of potential planets on the basis of disk morphology. In this section, we explore the 176 of these sources that are in \gaia\ DR3 to evaluate whether there is also evidence of massive planets from \gaia\ astrometry. Figure~\ref{fig:176_ruwe_period} shows \ruwe\ and the upper bound on orbital period of potential massive planets from \citetalias{pearce2022}. These limits on the orbital periods of the candidates are largely consistent with the sweet spot for potential impact on \ruwe. This coincidence motivated this work.

\begin{figure}
    \centering
    \includegraphics[width=1.0\linewidth]{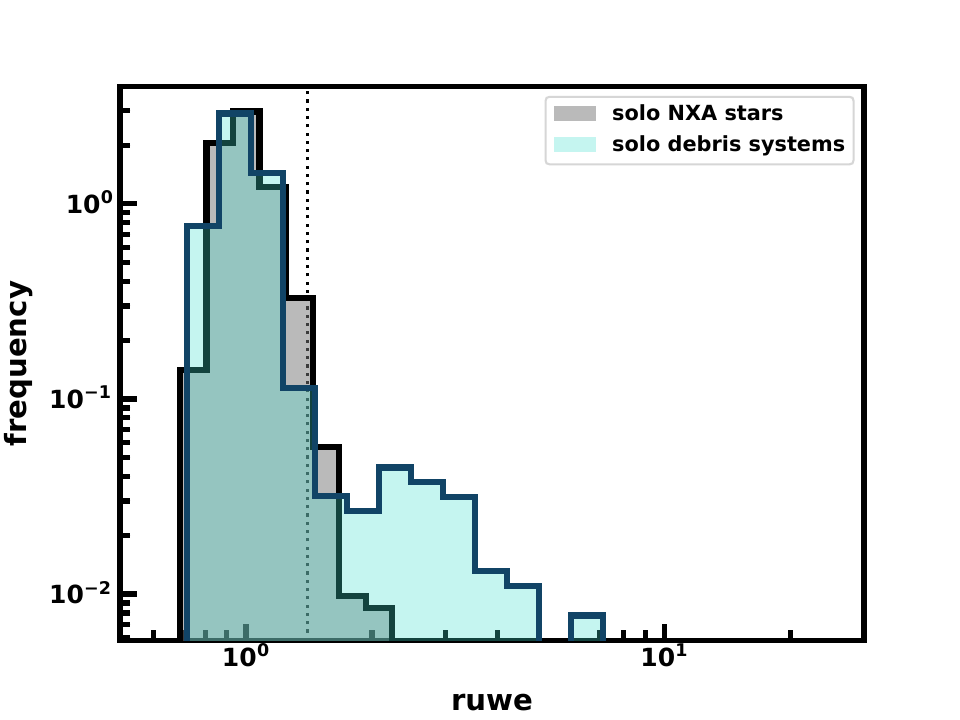}
    \caption{Histogram indicating the distribution of \ruwe\ values in the debris system stars. The blue-green histogram refers to these stars, all of which show no evidence of a stellar companion, as described in the text. The gray histogram corresponds to solo NXA stars. The excess source count above the $\ruwe = 1.4$ line suggests that the debris-disk systems contain promising candidates for planetary hosts.}\label{fig:176_ruwe_hist}
\end{figure}

To hone in on planetary companions, we check whether sources are associated with known stellar binaries using SIMBAD \otypes\ and \gaia's \nss\ flag. We identify 55 potential stellar binaries. The remaining 121 stars, ``solo'' debris-disk systems, have \ruwe\ distributed as in Figure~\ref{fig:176_ruwe_hist}. Eleven sources (9.1\%) have \ruwe\ above 1.4, which is a significant excess compared with the solo NXA stars (proportion-$z$ test $p$-value of 0.003). 

Among the \citetalias{pearce2022} candidates, 15 were known planet hosts. Since then, two other planetary systems have been discovered in the list of 178: AF~Lep (HD 35850), and HD~114082. We also count 
HN Peg as a planet host though \citetalias{pearce2022} do not. (HN Peg b is listed in the \nxa\ as a 20~\mjup\ body, but it is well outside of the debris disk around the host star.)   Five of the 18 planet hosts have massive planets in the sweet spot of orbital periods. These five, and two others (not in the sweet spot) around ``solo'' stars, are included in the ML training set described in \S\ref{sec:ml}. One source in the debris-disk systems list, \mbox{$\tau$ Cet}, is listed in the NXA (as obtained from IPAC through the Python \texttt{astroquery.ipac} module), with multiple planetary candidates identified with radial velocity measurements \citep{tuomi2013, feng2017}. Since the \citetalias{pearce2022} analysis, follow-up observations have questioned the earlier detections \citep{korolik2023, harada2025}. We nonetheless include \mbox{$\tau$ Cet} as a ``planet host,'' but do not incorporate it in our machine-learning algorithm's training set.  

Table~\ref{tab:176topruwe} lists the top 25 individual debris-disk systems ranked by high \ruwe\ values. All have bright hosts, with $G = 2$--6. Because detector saturation or stray light may affect astrometry, we check consistency between brightness in G compared to blue ($B_P$) and red ($R_P$) bands using the color excess factor $C_\ast$, defined in \citet[Eq.~6 and Table~2 therein]{riello2021}. The top four brightest stars ($\alpha$~CrB, $\beta$~UMa, $\gamma$~Boo and $\tau$~Cet) in the table have anomalous color excess ($|C_\ast| > 0.05$), perhaps from blending (three of these sources are binaries) or detector saturation ($\photgmeanmag \leq 3.3$~mag). We exclude these sources from further analysis as planet hosts. All other stars for which we have \gaia\ astrometry (including those not listed in the table) show no significant color excess. 

Sometimes, a high proper motion impacts the overall quality of astrometry. Five sources, including two in Table~\ref{tab:176topruwe} ($\tau$~Cet and $\epsilon$~Eri), have proper motions close to or over 1,000~mas/yr. Of these five systems, only one (BD-07~4003), with $\ruwe = 1.303$, is not flagged as a possible binary in Gaia (through the \nss\ flag) or SIMBAD (``**'' in the \otypes\ list). In the case of \mbox{$\tau$ Cet}, SIMBAD's \otypes\ contain **, as it may be in an optical binary, not a physical one. (Here, to be prudent in our selection of truly single stars, we do not attempt to determine the reason a source has a double-star designation in SIMBAD.)

Alongside \ruwe, the machine learning classifier from \S\ref{sec:ml} provides a separate measure of the probability of a planetary companion, $\probpl$. Table~\ref{tab:176topml} lists the top 25 most probable candidate planet hosts. As in Table~\ref{tab:176topruwe}, this table includes columns indicating known companions. Five of the top 25 sources have confirmed massive planets in the sweet spot of orbital periods. They are the only debris systems that were included in the ML classifier training set. The classifier is designed to select these systems as planet hosts and, as expected, all appear in the table with $\probpl > 80$\%. Three of these systems have the highest probability of being planet hosts (the top three rows in Table~\ref{tab:176topml}).

\begin{deluxetable*}{lrrrlrcrrr}
\tabletypesize{\footnotesize}
\tablecaption{Debris systems with the highest \ruwe\ values
\label{tab:176topruwe}}
\tablehead{
  \colhead{SIMBAD} & 
  \colhead{$\varpi$} & 
  \colhead{$\mu$} & 
  \colhead{$G$} & 
  \colhead{$B_P$-$R_P$} & 
  \colhead{\ } & 
  \colhead{companions} & 
  \colhead{$\mmin$} & 
  \colhead{$T_\text{max}$} 
  \\
  \colhead{\texttt{main\_id}} & 
  \colhead{\scriptsize{(mas)}} & 
  \colhead{\scriptsize{(mas/yr)}} & 
  \colhead{\scriptsize{(mag)}} & 
  \colhead{\scriptsize{(mag)}} & 
  \colhead{\ruwe} & 
  \colhead{\scriptsize{stars\ \ planets}} & 
  \colhead{\scriptsize{(\mjup)}} & 
  \colhead{\scriptsize{(yr)}} & 
  \colhead{$\probpl$}
  }
\startdata 
\rowcolor{gray!15}
V* DE Boo    &    85.993 & 491.670 &  5.766 & \ 1.041 & 16.496 & 1 \ \ \ \ 0 &  0.20 & 750.07 & 0.61 \\ 
\rowcolor{gray!15}
* alf CrB    &    42.241 & 147.773 &  2.269 & \ 0.530\ x & 10.338 & 1 \ \ \ \ 0 &  0.09 &  3.20 & 0.67 \\ 
* bet UMa    &    38.603 & 86.260 &  2.399 & \ 0.494\ x &  5.991 & 0 \ \ \ \ 0 &  0.20 & 55.58 & 0.58 \\ 
* zet Lep    &    44.794 & 14.224 &  3.541 & \ 0.244 &  4.317 & 0 \ \ \ \ 0 &  0.24 & 53.14 & 0.75 \\ 
* eta Lep    &    66.857 & 145.268 &  3.636 & \ 0.539 &  4.160 & 0 \ \ \ \ 0 &  0.08 & 48.75 & 0.79 \\ 
*  10 Tau    &    71.837 & 534.697 &  4.141 & \ 0.758 &  3.063 & 0 \ \ \ \ 0 &  0.03 & 16.24 & 0.59 \\ 
* eps Pav    &    31.259 & 154.863 &  3.963 & -0.017 &  3.007 & 0 \ \ \ \ 0 &  1.20 & 107.41 & 0.61 \\ 
* gam Oph    &    33.617 & 78.214 &  3.753 & \ 0.109 &  2.879 & 0 \ \ \ \ 0 &  0.60 & 298.56 & 0.60 \\ 
\rowcolor{gray!15}
* eps Eri    &   310.577 & 974.982 &  3.466 & \ 1.140 &  2.717 & 1 \ \ \ \ 1 &  0.19 & 422.47 & 0.55 \\ 
\rowcolor{gray!15}
* iot Psc    &    73.237 & 576.636 &  3.999 & \ 0.705 &  2.640 & 1 \ \ \ \ 0 &  0.07 & 151.26 & 0.63 \\ 
\rowcolor{gray!15}
* tau Cet    &   273.810 & 1922.319 &  3.300 & \ 1.051\ x &  2.634 & 1 \ \ \ \ 4 &  0.01 & 16.86 & 0.50 \\ 
* kap Phe    &    41.723 & 111.244 &  3.919 & \ 0.262 &  2.610 & 0 \ \ \ \ 0 &  0.24 & 119.08 & 0.83 \\ 
\rowcolor{gray!15}
* gam Boo    &    37.913 & 190.037 &  3.022 & \ 0.480\ x &  2.589 & 1 \ \ \ \ 0 &  0.60 & 779.57 & 0.55 \\ 
* lam Boo    &    32.588 & 246.191 &  4.169 & \ 0.142 &  2.445 & 0 \ \ \ \ 0 &  0.10 & 11.69 & 0.53 \\ 
*  30 Mon    &    25.344 & 70.739 &  3.902 & \ 0.011 &  2.155 & 0 \ \ \ \ 0 &  0.40 & 58.72 & 0.60 \\ 
* gam Dor    &    48.949 & 209.816 &  4.188 & \ 0.455 &  2.008 & 0 \ \ \ \ 0 &  0.07 & 34.17 & 0.76 \\ 
\rowcolor{gray!15}
* sig And    &    23.254 & 77.687 &  4.502 & \ 0.095 &  1.971 & 1 \ \ \ \ 0 &  0.30 & 116.19 & 0.54 \\ 
\rowcolor{gray!15}
* bet Cir    &    33.820 & 167.199 &  4.061 & \ 0.137 &  1.910 & 1 \ \ \ \ 0 &  0.21 & 53.21 & 0.54 \\ 
\rowcolor{gray!15}
* eta Crv    &    54.813 & 428.573 &  4.210 & \ 0.526 &  1.897 & 1 \ \ \ \ 0 &  0.34 & 899.56 & 0.75 \\ 
* d Sco      &    23.539 & 106.294 &  4.782 & \ 0.026 &  1.619 & 0 \ \ \ \ 0 &  0.40 & 85.05 & 0.59 \\ 
\rowcolor{gray!15}
* pi.01 Ori  &    27.007 & 134.668 &  4.636 & \ 0.128 &  1.477 & 1 \ \ \ \ 0 &  0.50 & 205.87 & 0.54 \\ 
\rowcolor{gray!15}
HD 218396    &    24.462 & 119.287 &  5.911 & \ 0.394 &  1.474 & 1 \ \ \ \ 4 &  2.50 & 1042.46 & 0.59 \\ 
* nu. Phe    &    65.527 & 688.511 &  4.828 & \ 0.728 &  1.381 & 0 \ \ \ \ 0 &  0.08 & 146.97 & 0.63 \\ 
HD 146897    &     7.565 & 28.175 &  8.976 & \ 0.678 &  1.379 & 0 \ \ \ \ 0 &  2.00 & 245.42 & 0.35 \\ 
\rowcolor{gray!15}
* mu. Cet    &    37.592 & 281.673 &  4.154 & \ 0.441 &  1.344 & 1 \ \ \ \ 0 &  0.30 & 451.84 & 0.71 
\enddata
\tablecomments{The first column is the main source identifier in SIMBAD, while the second column (\texttt{parallax}) through the sixth (\ruwe) are from \gaia\ DR3. In the color index column (\bprp), an 'x' marks sources with a color excess factor $|C_\ast| > 0.05$, suggesting an anomaly \citep{riello2021}. The ``stars'' and ``planets'' columns indicate companions. Rows with known companions have a gray background; the exception, also in gray, is $\tau$~Cet, which is likely an optical, not physical binary (despite the ** indicator in SIMBAD), and its planets (listed in the NXA) are currently unconfirmed. The predicted minimum planet mass ($\mmin$) and maximum orbital period ($\permax$) are from \citetalias{pearce2022}, while $\probpl$ is our ML planet-host likelihood.
\vspace{-22pt}
}
\end{deluxetable*}
\begin{deluxetable*}{lrrrlrcrrr}
\tabletypesize{\footnotesize}
\tablecaption{Debris systems highlighted by machine learning
\label{tab:176topml}}
\tablehead{
  \colhead{SIMBAD} & 
  \colhead{$\varpi$} & 
  \colhead{$\mu$} & 
  \colhead{$G$} & 
  \colhead{$B_P$-$R_P$} & 
  \colhead{\ } & 
  \colhead{companions} & 
  \colhead{$\mmin$} & 
  \colhead{$T_\text{max}$} & 
    \colhead{\ }
  \\
  \colhead{\texttt{main\_id}} & 
  \colhead{\scriptsize{(mas)}} & 
  \colhead{\scriptsize{(mas/yr)}} & 
  \colhead{\scriptsize{(mag)}} & 
  \colhead{\scriptsize{(mag)}} & 
  \colhead{\ruwe} & 
  \colhead{\scriptsize{stars\ \ planets}} & 
  \colhead{\scriptsize{(\mjup)}} & 
  \colhead{\scriptsize{(yr)}} &
  \colhead{$\probpl$}
  }
\startdata
\rowcolor{gray!15}
HD  50554    &    32.185 & 103.098 &  6.715 & \ 0.729 &  1.002 & 0 \ \ \ \ 1 &  0.08 & 248.06 & 0.99 \\ 
\rowcolor{gray!15}
HD 114082    &    10.520 & 44.498 &  8.106 & \ 0.568 &  0.944 & 0 \ \ \ \ 1 &  1.10 & 85.11 & 0.99 \\ 
\rowcolor{gray!15}
* q01 Eri    &    57.641 & 196.720 &  5.383 & \ 0.711 &  0.875 & 0 \ \ \ \ 1 &  0.16 & 332.59 & 0.97 \\ 
V* V1358 Ori &    19.353 & 43.465 &  7.707 & \ 0.724 &  0.930 & 0 \ \ \ \ 0 &  0.20 & 10.08 & 0.91 \\ 
\rowcolor{gray!15}
* tau01 Gru  &    30.192 & 232.513 &  5.887 & \ 0.772 &  1.004 & 0 \ \ \ \ 1 &  0.08 & 220.19 & 0.89 \\ 
V* DK Cet    &    24.205 & 115.255 &  7.914 & \ 0.835 &  0.967 & 0 \ \ \ \ 0 &  0.32 & 40.76 & 0.84 \\ 
* kap Phe    &    41.723 & 111.244 &  3.919 & \ 0.262 &  2.610 & 0 \ \ \ \ 0 &  0.24 & 119.08 & 0.83 \\ 
HD 127821    &    31.558 & 177.419 &  5.998 & \ 0.574 &  0.974 & 0 \ \ \ \ 0 &  0.21 & 379.39 & 0.83 \\ 
HD 209253    &    31.804 & 30.261 &  6.505 & \ 0.668 &  1.103 & 0 \ \ \ \ 0 &  0.30 & 439.14 & 0.82 \\ 
\rowcolor{gray!15}
V* AF Lep    &    37.254 & 52.138 &  6.210 & \ 0.736 &  0.918 & 0 \ \ \ \ 1 &  1.10 & 184.46 & 0.80 \\ 
* eta Lep    &    66.857 & 145.268 &  3.636 & \ 0.539 &  4.160 & 0 \ \ \ \ 0 &  0.08 & 48.75 & 0.79 \\ 
HD  38397    &    18.659 & 29.129 &  7.999 & \ 0.737 &  1.062 & 0 \ \ \ \ 0 &  0.60 & 154.57 & 0.78 \\ 
HD 107146    &    36.404 & 229.612 &  6.904 & \ 0.793 &  0.993 & 0 \ \ \ \ 0 &  0.44 & 264.37 & 0.77 \\ 
HD   8907    &    29.941 & 111.074 &  6.537 & \ 0.660 &  1.058 & 0 \ \ \ \ 0 &  0.28 & 289.66 & 0.77 \\ 
* gam Dor    &    48.949 & 209.816 &  4.188 & \ 0.455 &  2.008 & 0 \ \ \ \ 0 &  0.07 & 34.17 & 0.76 \\ 
*   8 Dra    &    33.967 & 30.347 &  5.145 & \ 0.429 &  1.137 & 0 \ \ \ \ 0 &  0.22 & 293.60 & 0.75 \\ 
* zet Lep    &    44.794 & 14.224 &  3.541 & \ 0.244 &  4.317 & 0 \ \ \ \ 0 &  0.24 & 53.14 & 0.75 \\ 
\rowcolor{gray!15}
* eta Crv    &    54.813 & 428.573 &  4.210 & \ 0.526 &  1.897 & 1 \ \ \ \ 0 &  0.34 & 899.56 & 0.75 \\ 
V* NZ Lup    &    16.627 & 78.557 &  7.794 & \ 0.834 &  0.902 & 0 \ \ \ \ 0 &  1.30 & 168.21 & 0.73 \\ 
HD 218340    &    17.815 & 121.364 &  8.295 & \ 0.787 &  1.091 & 0 \ \ \ \ 0 &  0.40 & 1259.06 & 0.73 \\ 
HD  61005    &    27.434 & 92.487 &  8.017 & \ 0.911 &  0.940 & 0 \ \ \ \ 0 &  0.70 & 273.70 & 0.71 \\ 
\rowcolor{gray!15}
* mu. Cet    &    37.592 & 281.673 &  4.154 & \ 0.441 &  1.344 & 1 \ \ \ \ 0 &  0.30 & 451.84 & 0.71 \\ 
\rowcolor{gray!15}
V* IS Eri    &    25.825 & 144.096 &  8.300 & \ 0.931 &  0.951 & 1 \ \ \ \ 0 &  0.15 & 37.63 & 0.69 \\ 
HD  48370    &    27.815 & 72.562 &  7.746 & \ 0.874 &  1.112 & 0 \ \ \ \ 0 &  0.40 & 91.28 & 0.69 \\ 
\rowcolor{gray!15}
* alf CrB    &    42.241 & 147.773 &  2.269 & \ 0.530\ x & 10.338 & 1 \ \ \ \ 0 &  0.09 &  3.20 & 0.67
\enddata
\tablecomments{The columns are the same as in Table~\ref{tab:176topruwe}, though sorted by $\probpl$, the machine-learning predicted likelihood that the source hosts a massive planet in the sweet spot of orbital periods. Rows with gray backgrounds indicate sources that are either in known stellar binaries or have known planets and were included in the ML training set.}
\end{deluxetable*}

Just over half of the list of high-\ruwe\ stars in Table~\ref{tab:176topruwe} are double stars. As a flag for binarity, \ruwe\ is playing its part in the table.  The list also contains three known planet hosts.  The remainder, including six of the top ten in \ruwe, are putatively single stars that are good candidates as substellar or planetary hosts. Table~\ref{tab:176topml}, with the top ML probabilities for planet hosts, contains many fewer double stars (4 of 25). It lists five planet hosts, including the top three entries, though only because these sources were in the ML training set. Seven stars are in both top-25 tables, four of which have no known binary partner. We consider these sources below. 

In both tables, it is striking that the minimum planet masses from the \citetalias{pearce2022} analysis are mostly less than one Jupiter mass.  As in \S\ref{sec:bg}, the detection of a planetary companion with astrometry typically requires a larger planetary mass. For the few candidates of interest that we consider next, we assume that the putative planets are indeed massive enough to impact the astrometry of their host star. Indeed, \citetalias[Fig.~7 therein]{pearce2022} indicate that the actual planetary mass may be a factor of three higher than the minimum mass in some systems. Still, as we mention below, at the population level, it is unlikely that most of the 178 debris-disk systems host a planet more massive than Jupiter \citep{yee2025}.

\subsection{Candidates of interest}

From Tables~\ref{tab:176topruwe} and \ref{tab:176topml}, we identify debris-disk systems as promising potential planet hosts. For each candidate, we indicate predicted limits on planetary mass and orbital distance or period. These values represent best estimates, as in \citetalias{pearce2022}, without consideration of quoted uncertainties. They also imply a correlation between specific values of planetary mass and orbital distance; the larger the assumed planetary mass, the smaller the planet's orbital distance \citep[\S{3.1} therein]{pearce2022}. We start with stars that appear in both tables.

\mysubsubsection{High \ruwe\ and high ML planet-host probability}

Four stars that appear in both top-25 tables and are neither known binaries nor known planet hosts are:

\begin{itemize}

\item{$\zeta$~Leporis} (HD  38678) is a bright (G = 3.5) A2~IV-V(n) star with \ruwe\ = 4.31. The predicted planet has a mass $\gtrsim$ 0.24~\mjup\ and orbital period $\lesssim$ 53 years. This orbit is well within the sweet spot, but for an astrometric detection, the planet mass would be $\gtrsim 1~\mjup$.

\item{$\eta$~Leporis} (HD 40136) is a bright (G = 3.63) F2~V star with a predicted planet (minimum mass $\mmin 0.08$~\mjup) in the sweet spot \citep[orbital period $\lesssim$ 50~years;][]{pearce2022}. Although its brightness may be a factor in its large \ruwe\ (4.2), a planetary companion with a mass of at least an order of magnitude larger than $\mmin$ could be sculpting the $\eta$~Lep debris disk. 

\item{$\kappa$~Phe} (HD 2262) is an A5~IV(n) star with G = 3.92 and \ruwe\ = 2.6. The predicted planet mass is $\gtrsim$ 0.24~\mjup\ with an orbital period $\lesssim$ 100 years. Though the predicted maximum orbital period lies outside the sweet spot, \citetalias{pearce2022} note that if the planet had a larger mass, then it would reside closer to the star.

\item{$\gamma$~Doradus} (HD 27290) is an F1~V star with G = 4.19 and $\ruwe = 2.0$. \citetalias{pearce2022} predict a mass $\gtrsim$ 0.1~\mjup\ and an orbital period $\lesssim$ 34 years. They suggest that there is flexibility in these estimates as the constraints on disk geometry are not tight.

\end{itemize}

Of these sources, the most promising candidate from an astrometric detection standpoint is \mbox{$\zeta$ Lep}; its predicted planet has the largest minimum mass of about 0.25~\mjup\ and a maximum orbital period of 53~years. The predictions for the other stars include either a smaller minimum mass or a larger orbital distance. In any case, all four are worthy of a time-series analysis in \gaia\ DR4 to investigate the source of their unusually high \ruwe\ values.

\mysubsubsection{High-\ruwe\ candidates with low ML $\probpl$} 

Other candidates of interest are stars with no known companion, high \ruwe ($>1.4$), a predicted mass $\gtrsim 0.1$~\mjup, and a maximum orbital period in the sweet spot, though with low ML planet-host probability ($<67$\%):

\begin{itemize}

\item{10~Tauri} (HD  22484) is a bright, single F9~IV-V star with G = 4.14 and \ruwe = 3.063, but it has a slightly lower ML probability of being a planet host (59\%). The \citetalias{pearce2022} prediction is a planet with $\gtrsim$10~Earth masses and an orbital period of $\lesssim$16 years. 

\item{30 Monocerotis} (HD  71155) is a bright (G = 3.90), single A0~Va star with \ruwe\ of 2.155. The predicted planet has $\mmin \sim 0.5~\mjup$ and $\permax \sim 60$~yr. 30~Mon is one of the higher-mass stars in the debris-disk systems catalog, which may make it more challenging for astrometric detections of planets. 

\item{d Scorpii} (HD 146624) is another single, intrinsically luminous A1~Va star with G = 4.78, \ruwe\ =1.619, a predicted planet mass $\gtrsim$0.5~\mjup\, and an orbital period of $\lesssim$85 years. Like 10~Tau and 30~Mon, this source did not make both top-25 lists because its ML probability score for planet hosting is 0.59, below the list cut-off.
 
\end{itemize}

\mysubsubsection{High-probability ML candidates with low \ruwe}

The machine-learning classification in Table~\ref{tab:176topml} highlights sources with features that include demographics and astrometric indicators of companions other than \ruwe. Here is a sample of stars with no known companions, high ML planet-host probability ($>67$\%), low \ruwe ($< 1.4$), and predicted orbits from \citetalias{pearce2022} in the sweet spot: 

\begin{itemize}

\item{V1358 Orionis} (HD 43989) is a G0~V star (G = 7.71) with a low \ruwe\ (0.930). \citetalias{pearce2022} infer a 0.3~\mjup\ (or greater) planet on an orbit with a period of 10~years (or less) as the source of debris disk sculpting. 

\item{DK Ceti} (HD 12039) is a somewhat cooler G4~V star with G = 7.91, \ruwe\ = 0.967, and predicted mass $\mmin \lesssim 0.2$~\mjup, similar to \mbox{V1358 Orionis}, but with an orbital period of $\permax \lesssim 30$~yr.

\item{HD 48370} is an even cooler G8--K0~V star with G = 7.75, \ruwe\ = 1.1, $\mmin \gtrsim 0.4$~\mjup, and $\permax \lesssim 90$~yr. 
\end{itemize}

Among all the candidates of interest listed above, most have estimated ages ranging from $\sim$200~Myr up to $\sim$5~Gyr \citepalias{pearce2022}. However, V1358~Ori ($\sim$40~Myr), DK~Cet ($\sim$45~Myr), and HD~48370 ($\sim$60~Myr) are much younger, have large 70--100~$\mu$m excess, and are members of nearby moving groups \citep[see also][]{gagne2026}. Approximately 6\% of {\it Spitzer} and {\it Herschel} debris-disk targets belong to a moving group \citep{kbn2026}; young systems with such large cold dust luminosities merit special attention once \gaia\ DR4 is released.

\mysubsubsection{Other noteworthy systems}

Two known planet hosts not part of the ML training set (they were both listed in SIMBAD as potential double stars) provide an assessment of the ML analysis:

\begin{itemize}

\item{$\epsilon$~Eri} (HD 22049) is a K2~V star (G = 3.47) that hosts a 0.7~\mjup\ planet in the sweet spot: a period of 7~years \citep{hatzes2000} and a semimajor axis of roughly 3~\au. The system contains warm dust at 3--20~au \citep{su2017, booth2023} and cold dust at $\sim$ 70~au \citep{beichman1985}.  \citetalias{pearce2022} propose a second planet with a mass of at least 0.2~\mjup\ near or within 50~au to account for stirring of the cold disk. SIMBAD lists this star as a double, though the \nxa\ shows no bound stellar companion, leaving the known planet as the potential source of $\epsilon$~Eri's high \ruwe\ of 2.7.

\item{HR 8799} (HD 218396), an F-type $\gamma$ Doradus pulsating variable, hosts four known giant planets and a broad cold debris disk centered at about 200~au \citep{marois2008}. The innermost planet with a mass of $\sim$ 10~\mjup\ has an orbital period ($\sim 50$~years), well inside the sweet spot \citep{marois2010}. It is reassuring that this system was flagged as a planet host both by its \ruwe\ value (1.474) and the ML classification (59\%).

\item{AF Lep}, which was used in our training set, has a single planet, \mbox{AF Lep b}, that was discovered \textit{after} \citetalias{pearce2022} completed their analysis \citep{mesa2023, derosa2023, franson2023}.  Their predictions ($\mmin \sim 1.1$~\mjup\ and $\permax \sim 200$~yr) are formally consistent with the observations \citep[mass $\sim$ 3~\mjup\ and orbital period $\sim$ 25~yr][]{mesa2023}. Still, the difference between the limits and observed parameters suggests that the planet is probably not responsible for features in the cold debris disk at 40--50~au \citep[cf.][Eq.~(6) therein]{pearce2014}.

\end{itemize}

The analysis in this section highlights individual debris-disk systems as promising candidates for future study. Certainly, the high \ruwe\ ($\gtrsim$1.4) sources are intriguing, as our analysis of planet hosts in \S\ref{sec:knowncomps} illustrates. That analysis also issues a warning: the majority of hosts that harbor jovian or super-jovian planets (Fig.~\ref{fig:176_ruwe_hist}), do not have suggestively high \ruwe\ values. Furthermore, our top candidates identified by \ruwe\ and by the ML algorithm have predicted minimum planetary masses that are mostly below 1~\mjup, roughly the mass needed to be detectable with \gaia\  DR3 astrometry (\S\ref{sec:bg}). For our candidates to be promising planet hosts, the predicted masses must be significant underestimates, or perhaps other unseen planetary or stellar companions are implicated.

\section{Discussion}\label{sec:discuss}

Our main goal is to find astrometric signatures of planets around stars for which there is other evidence of planetary companions. Because planets can stir and sculpt dusty debris, the systems discussed in \citetalias{pearce2022} offer an excellent opportunity to explore this possibility. We start with criteria for identifying companions in Gaia DR3 based on stellar binaries. We assess how these criteria fare as indicators of exoplanets using stellar hosts drawn from the NASA Exoplanet Archive. In both cases, we consider only sources that are similar to those stars in parallax, brightness, and color. While this process reduces the number counts substantially (only 755 of over 6,000 stellar hosts are similarly bright and close as the debris disk systems), it adds some assurance that measures of astrometric quality are not the result of selection effects.

We confirm a clear connection between the Gaia astrometric quality indicator \ruwe\ and the orbital period of known binary stars. For periods in a sweet spot between days and centuries, approximately spanning the duration of Gaia observations, binary motion impacts astrometric quality as expected for stars selected similarly to the debris-disk systems ({\S}2 and Fig.~\ref{fig:ruwe} above; see also \citealt{gaiabin2023}). Astrometric quality is not a perfect identifier of binary systems. Roughly 7\%\ of the binary stars in the sweet spot have $\ruwe < 1.4$; perhaps the geometry and orientation of their orbits fail to impact fitting with a linear drift model. Conversely, putative single stars may have high \ruwe\ values, although these may have unidentified massive companions. Despite these concerns, in a set of well-studied planet hosts with no known stellar or planetary companion in the sweet spot (the solo NXA stars considered here), fewer than 3\%\ have $\ruwe > 1.4$ (Table~\ref{tab:ruwestats}).

Because stellar reflex motion from planetary companions is typically much smaller than from stellar companions, astrometric quality is less effective for planet detection. Still, on a population level, significantly more stars that host massive planets with orbital periods in the sweet spot have higher \ruwe\ than the ``solo'' stars with no known massive companions in that orbital regime (Table~\ref{tab:plprobs}). The fraction of high-\ruwe\ stars with massive planets ($q \geq 10$~\mjup/\msun) is $\sim$ 16\%, compared with 3\%\ for solo stars. This correspondence between astrometry and planets has aided in identifying several planetary systems \citep[e.g.,][]{currie2020,currie2023,currie2026,stefansson2025,vioque2026}. Yet, the majority of planet hosts have low values of \ruwe, a reminder that detection of planetary companions can be a serious challenge with \gaia\ astrometry. 

With this background, we consider the 176 debris disk systems listed by \citetalias{pearce2022} that are also in the Gaia DR3 archive. Of the 121 sources with no indicator of a stellar companion, 11 have \ruwe\ above the 1.4 threshold. As with the NXA stellar hosts, this distribution suggests a population with massive planets. On a source level, we identify several promising candidates on the basis of their excess \ruwe\ values and the high probability assigned by our machine-learning analysis that they are planet hosts. Since the machine learning algorithm mixed stellar demographics with the astrometry, it provides a separate view of potential hosts than \ruwe\ alone.

With lower limits for the masses of predicted planets in \citetalias{pearce2022}, it is tempting to consider the impact of larger planet masses closer to the host inside the debris disks in these systems. However, the measured mass functions of planets from direct imaging, microlensing, and radial velocity measurements place strong constraints on planet candidates \citep{yee2025}. In particular, making \textit{all} of these candidates significantly more massive creates a strong tension with the microlensing mass function, which in turn roughly agrees with the mass functions of directly imaged and radial velocity planets \citep{fulton2021,vigan2021,yee2025}. While several of the debris disk planet candidates could be more massive than the \citetalias{pearce2022} lower limits, it is unlikely that \textit{many} are more massive. Thus, the highest probability systems derived here are the most likely planet candidates in the \citetalias{pearce2022} sample, based on \ruwe\ and the ML analysis. 

To compare these results with giant planet frequencies derived from direct imaging and radial velocity studies, we estimate the implied frequency of giant planets. Some stars in the \citetalias{pearce2022} sample are A-type stars, where the frequency of debris disks is $\sim$ 33\% \citep[e.g.,][]{su2006}. Most are FGK stars, which have a somewhat lower debris disk frequency $\sim$ 20\% \citep[e.g.][]{bryden2009, carp2009a, eiroa2013, sibthorpe2018}. If we adopt 25\% as the intrinsic debris disk frequency of the mix of stars in \citetalias{pearce2022}, the 10\% frequency of \citetalias{pearce2022} single stars with $\ruwe > 1.4$ suggests an occurrence rate of $\sim$ 2\%--3\% for giant planets in the sweet spot of \gaia\ orbital periods. This estimate compares well with direct imaging and radial velocity estimates: \citet[][their Figure 5]{fernandes2019} report a 1--2\%\ occurrence rate of 1--20~\mjup\ planets at 10--100~\au. Including smaller orbital distances ($<10$~\au) allows larger rates, 5--10\%, for Jupiter mass planets \citep[e.g.,][and Table 2, above]{fulton2021,vigan2021}.

\section{Conclusion}\label{sec:conclude}

This work is an exploration of the potential and limitations of astrometry for planet discovery. Here, we study 176 bright, nearby stars that host debris disks to seek astrometric evidence of planets in these systems from \gaia\ DR3. A review of how stellar and planetary companions might impact \gaia\ astrometry in theory is provided in \S\ref{sec:bg}. We test the analytic results using known binaries from multiple catalogs and also planet hosts from the NASA Exoplanetary Archive. Because of their low mass compared with their host star, even giant planets can only marginally affect astrometry in Gaia DR3 (\S\ref{sec:bg}). Still, single stars with known planets at optimal orbital periods for detection by Gaia (the ``sweet spot'' considered above) show at a population level that Gaia's astrometric quality parameter \ruwe\ is sensitive to the presence of planets (Fig.~\ref{fig:nxa_qpl_ruwe}).

We follow this promising result with an analysis of the debris-disk systems. As metrics for planet detection, we focus on \ruwe\ and a planet-hosting probability, $\probpl$, from a machine-learning algorithm (\S\ref{sec:knowncomps}).  Our main results, including a comparison with the predictions of \citetalias{pearce2022} for limits on mass and orbital distance of planets that could stir and sculpt the dusty debris, are as follows:

\begin{itemize} 

\item The 176 debris-disk systems are a mix of single stars (118 objects), binary systems (58), and known planet hosts (18). A subset of 113 stars with no known planetary or stellar companions (``solo'' stars) shows significantly higher \ruwe\ values than similarly selected planet hosts with no known stellar companions or planets more massive than Saturn (the solo NXA stars).

\item Among the solo debris-disk systems, a dozen have $\ruwe > 1.4$ (7\% of the total), and ten systems have  $\ruwe > 2$ (\S\ref{sec:176} and Fig.\ \ref{fig:176_ruwe_hist}). For comparison, less than 3\%\ of the solo NXA stars have $\ruwe > 1.4$; none have $\ruwe > 2$ (Fig.~\ref{fig:nxa_qpl_ruwe}). We recommend follow-up studies of a dozen systems to identify new stellar, substellar or planetary companions.

\item The machine-learning algorithm in \S\ref{sec:ml} complements the analysis based on \ruwe. Trained on single planet hosts with and without giant planets, the model weights a mix of astrometric and photometric features to determine the probability that a star is a planet host. The strategy includes a broader suite of parameters than \ruwe\ that may inform if and how a planet and the reflex motion it induces impact \gaia's measurements.

\item Sixteen single debris-disk systems with no known planets were flagged as promising candidates by the machine-learning classifier ($\probpl >67$\%; Table~\ref{tab:176topml}).  Four of these systems have $\ruwe > 1.4$. We caution that the ML classifier may respond to selection biases in planet detection methods. A recommendation is to track the high $\probpl$ sources in future studies that might reveal as-yet-undetected companions as a way to validate this general approach.

\item Overall, we found little connection between \ruwe, $\probpl$, and $\mmin$, the predicted minimum planet mass from \citetalias{pearce2022}. In the accompanying data files and in Tables~\ref{tab:176topruwe} and \ref{tab:176topml}, the predicted minimum mass is generally low; only two of the two dozen solo debris-disk systems in either the top~25 \ruwe\ stars or the top~25 ML-derived list have minimum mass $\mmin > 1$~\mjup, and both of these have a maximum orbital period $T_\text{max}$ greater than a century, outside the sweet spot for \gaia\ astrometry.

\item Associating our highest ranked planet-host candidates with the predicted planets from \citetalias{pearce2022} requires adopting significantly larger planetary masses. Since more massive planets can stir and sculpt dust at greater separation, they could be closer to the host star, increasing their astrometric detectability.

\item Alternatively, the most promising candidates reported here may harbor unrelated massive planets close to their stellar hosts. While postulating the existence of high-mass planets may be reasonable for some stars in the \citetalias{pearce2022} sample, planet mass limits from microlensing \citep[e.g.,][]{yee2025} mean that adopting large masses for the whole population is not realistic. 

\item Other explanations for the lack of association between predicted mass limits from \citetalias{pearce2022} and our top planet-host candidates include unknown stellar companions (high-\ruwe), or spurious detections (\ruwe\ statistical outliers or if the ML probabilities are weighted too heavily by photometry, not astrometry).

\item The number of sources used directly in this study is under a thousand. A final recommendation for future work is to broaden the selection criteria to consider the full NXA catalog of roughly 6,000 planet hosts to interpret \ruwe\ and train the machine-learning algorithm. Applying the ML model to more debris-disk systems \citep[e.g.,][and references therein]{kbn2026} might yield promising new candidates.

\end{itemize}

\vspace*{0.1pt}
While astrometric quality indicators like \ruwe\ are valuable flags of substellar or planetary companions, they are not designed to reveal reflex motion, the signature of a star with a massive companion. Adding in additional astrometric information by including measurements outside of the \gaia\ observing time frame, as in the Tycho-Gaia Astrometric Solution catalog \citep{tgas2015}, can extend the sensitivity for long-period planets. Yet \gaia\ alone has a wealth of data waiting in the wings. With its time-resolved astrometry, Gaia Data Release 4 will provide snapshots of stars, dots along their orbital paths \citep[e.g.,][]{brown2025, walczak2024}. With the release of DR4 just around the corner, we are eager to see how those dots connect.

\acknowledgements

We  thank anonymous referees for comments and suggestions that led to improvements in this work. The NASA Exoplanets Research Program, through contract 80NSSC24K0158, supported this project. EMP is grateful for support from the Undergraduate Research Opportunity Program at the University of Utah. This research has made use of data from the European Space Agency (ESA) mission \gaia\  (\url{https://www.cosmos.esa.int/gaia}), processed by the \gaia\ Data Processing and Analysis Consortium (DPAC, \url{https://www.cosmos.esa.int/web/gaia/dpac/consortium}). Funding for the DPAC has been provided by national institutions, in particular the institutions participating in the \gaia\ Multilateral Agreement. This research has also made use of the SIMBAD database \citep{wenger2000} and the VizieR catalog access tool \citep{vizier2000}, operated at CDS, Strasbourg, France. We are thankful for the availability of these resources. This project was supported by the NASA Exoplanets Research Program through contract 80NSSC24K0158.

\bibliography{main}
\bibliographystyle{aasjournal}

\end{document}